\documentclass{sig-alternate-ipsn13}

\usepackage{fixacm} 
\usepackage[small]{caption}
\usepackage{times}

\usepackage{graphicx}
\graphicspath{{figs/}}
\usepackage{paralist}

\usepackage{todonotes}  
\usepackage{color}

\usepackage{epstopdf}
\usepackage{subfigure}
\usepackage{amssymb}
\usepackage{amsfonts}
\usepackage{bm} 
\usepackage{mathtools}  
\usepackage{cool}  
\usepackage{mleftright}  
\usepackage{booktabs}
\usepackage[per-mode=symbol]{siunitx}
\usepackage{url}

\usepackage[amsmath,thmmarks,thref]{ntheorem}
\usepackage{algorithm}
\usepackage{algpseudocode}  
\theoremstyle{definition}

\usepackage{truonglatexdefs}

\usepackage{enumitem, kantlipsum}

\usepackage[capitalise]{cleveref}

\usepackage{balance}

\usepackage{etoolbox}
\makeatletter
\patchcmd{\maketitle}{\@copyrightspace}{}{}{}
\makeatother

\apptocmd{\thebibliography}{\small}{}{}

\begin{document}
%
\title{Data-Driven Modeling, Control and Tools \\ for Cyber-Physical Energy Systems}


%
%
%
\numberofauthors{1} 
%
\author{
%
%
\alignauthor
\Large Madhur Behl, Achin Jain and Rahul Mangharam\\
       \affaddr{Electrical and Systems Engineering, University of Pennsylvania, Philadelphia, USA.}\\
       \email{\{mbehl, achinj, rahulm\}@seas.upenn.edu}
}

\maketitle

\begin{abstract}
Demand response (DR) is becoming increasingly important as the volatility on the grid continues to increase.
Current DR approaches are completely manual and rule-based or involve deriving first principles based models which are extremely cost and time prohibitive to build.
We consider the problem of data-driven end-user DR for large buildings which involves predicting the demand response baseline, evaluating fixed rule based DR strategies and synthesizing DR control actions.
We provide a model based control with regression trees algorithm (mbCRT), which allows us to perform closed-loop control for DR strategy synthesis for large commercial buildings. Our data-driven control synthesis algorithm outperforms rule-based DR by $17\%$ for a large DoE commercial reference building and leads to a curtailment of $380\si{\kilo\watt}$ and over $\$45,000$ in savings.
Our methods have been integrated into an open source tool called DR-Advisor, which acts as a recommender system for the building's facilities manager and provides suitable control actions to meet the desired load curtailment while maintaining operations and maximizing the economic reward.
DR-Advisor achieves $92.8\%$ to $98.9\%$ prediction accuracy for 8 buildings on Penn's campus. 
We compare DR-Advisor with other data driven methods and rank $2^{nd}$ on ASHRAE's benchmarking data-set for energy prediction. 
\end{abstract}

%

\section{Introduction}
\label{sec:intro}
In 2013, a report by the U.S. National Climate Assessment provided evidence that the most recent decade was the nation's warmest on record~\cite{melillo2014climate} and experts predict that temperatures are only going to rise.
In fact, the year 2015 is very likely to become the hottest year on record since the beginning of weather recording in 1880~\cite{noaa}. 
Heat waves in summer and polar vortexes in winter are growing longer and pose increasing challenges to an already over-stressed electric grid. 

Furthermore, with the increasing penetration of renewable generation, the electricity grid is also experiencing a shift from predictable and dispatchable electricity generation to variable and non-dispatchable generation. 
This adds another level of uncertainty and volatility to the electricity grid as the relative proportion of variable generation vs. traditional dispatchable generation increases.
The organized electricity markets across the world all use some variant of real-time price for wholesale electricity. 
The real-time electricity market at PJM, one of the world's largest independent system operator (ISO), is a spot market where electricity prices are calculated at five-minute intervals based on the grid operating conditions. 
The volatility due to the mismatch between electricity generation and supply  further leads to volatility in the wholesale price of electricity.
For \eg the polar vortex triggered extreme weather events in the U.S. in January 2014, which caused many electricity customers to experience increased costs.
Parts of the PJM electricity grid experienced a $86$ fold increase in the price of electricity from $\$31/\si{\mega\watt\hour}$ to $\$2,680/\si{\mega\watt\hour}$in a matter of a few minutes~\cite{volatility}. 
Similarly, the summer price spiked $32$ fold from an average of $\$25/\si{\mega\watt\hour}$ to $\$800/\si{\mega\watt\hour}$ in July of 2015.
Such events show how unforeseen and uncontrollable circumstances can greatly affect electricity prices that impact ISOs, suppliers, and customers. 
Energy industry experts are now considering the concept that extreme weather, more renewables and resulting electricity price volatility, could become the new norm.

Across the United States, electric utilities and ISOs are devoting increasing attention and resources to demand response (DR)~\cite{goldman2010coordination}. Demand response is considered as a reliable means of mitigating the uncertainty and volatility of renewable generation and extreme weather conditions and improving the grid's efficiency and reliability.
The potential demand response resource contribution from all U.S. demand response programs is estimated to be nearly 72,000 megawatts (MW), or about 9.2 percent of U.S. peak demand~\cite{federal2008assessment} making DR the largest virtual generator in the U.S. national grid.
The annual revenue to end-users from DR markets with PJM ISO alone is more than $\$700$ million~\cite{pjm}. 
Global DR revenue is expected to reach nearly $\$40$ billion from 2014 through 2023~\cite{navigant}.

The volatility in real-time electricity prices poses the biggest operational and financial risk for large scale end-users of electricity such as large commercial buildings, industries and institutions~\cite{Mulhall2014327}; often referred to as \textit{C/I/I} consumers. 
In order to shield themselves from the volatility and risk of high prices, such consumers must be more flexible in their electricity demand. 
Consequently, large \textit{C/I/I} customers are increasingly looking to demand response programs to help manage their electricity costs.

DR programs involve a voluntary response of a building to a price signal or a load curtailment request from the utility or the curtailment service provider (CSP). 
Upon successfully meeting the required curtailment level the end-users are financially rewarded, but may also incur penalties for under-performing and not meeting a required level of load curtailment.
On the surface demand response may seem simple. Reduce your power when asked to and get paid. 
However, in practice, one of the biggest challenges with end-user demand response for large scale consumers of electricity is the following: \emph{Upon receiving the notification for a DR event, what actions must the end-user take in order to achieve an adequate and a sustained DR curtailment?} 
\begin{figure*}
\centering
\includegraphics[width=0.7\linewidth]{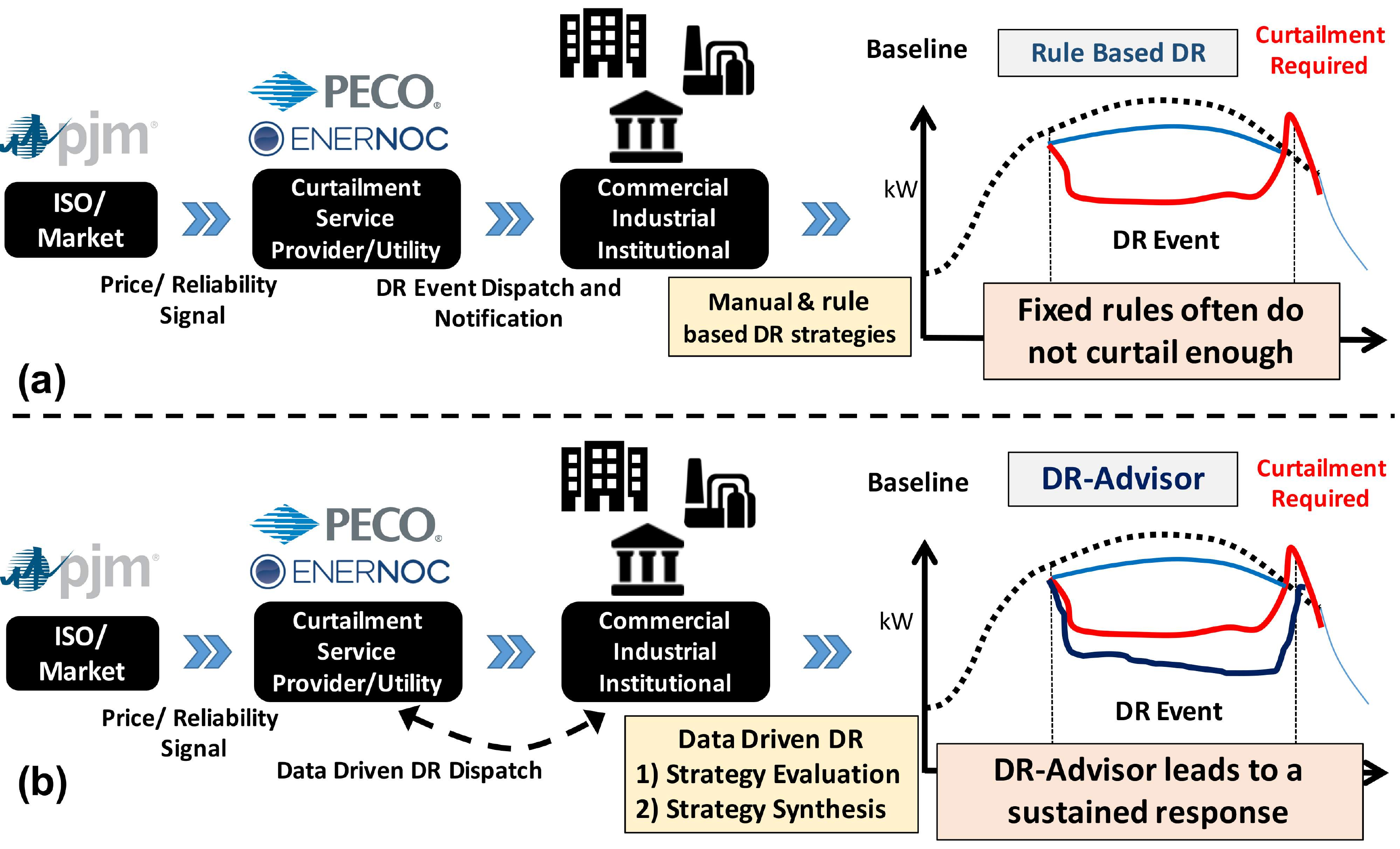}
\caption{Majority of DR today is manual and rule-based. (a) The fixed rule based DR is inconsistent and could under-perform compared to the required curtailment, resulting in DR penalties. (b) Using data-driven models DR-Advisor uses DR strategy evaluation and DR strategy synthesis for a sustained and sufficient curtailment.}
\label{fig:twocolumn}
\vspace{-3pt}
\end{figure*}
This is a hard question to answer because of the following reasons:
\begin{enumerate}[leftmargin=0.5cm,topsep=1pt,itemsep=-1ex,partopsep=1ex,parsep=1ex]
\item \textbf{Modeling complexity and heterogeneity}: Unlike the automobile or the aircraft industry, each building is designed and used in a different way and therefore, it must be uniquely modeled. 
Learning predictive models of building's dynamics using first principles based approaches (\eg with EnergyPlus~\cite{Crawley2001319}) is very cost and time prohibitive and requires retrofitting the building with several sensors~\cite{sturzeneggermodel};
The user expertise, time, and associated sensor costs required to develop a model of a single building is very high.
This is because usually a building modeling domain expert typically uses a software tool to create the geometry of a building from the building design and equipment layout plans, add detailed information about material properties, about equipment and operational schedules. 
There is always a gap between the modeled and the real building and the domain expert must then manually tune the model to match the measured data from the building~\cite{new2012autotune}. 
\item \textbf{Limitations of rule-based DR}: The building's operating conditions, internal thermal disturbances and environmental conditions must all be taken into account to make appropriate DR control decisions, which is not possible with using rule-based and pre-determined DR strategies since they do not account for the state of the building but are instead based on best practices and rules of thumb. As shown in Fig.~\ref{fig:twocolumn}(a), the performance of a rule-based DR strategy is inconsistent and can lead to reduced amount of curtailment which could result in penalties to the end-user. In our work, we show how a data-driven DR algorithm outperforms a rule-based strategy by $17\%$ while accounting for thermal comfort.
Rule based DR strategies have the advantage of being simple but they do not account for the state of the building and weather conditions during a DR event.
Despite this lack of predictability, rule-based DR strategies account for the majority of DR approaches.
\item \textbf{Control complexity and scalability}: Upon receiving a notification for a DR event, the building's facilities manager must determine an appropriate DR strategy to achieve the required load curtailment. 
These control strategies can include adjusting zone temperature set-points, supply air temperature and chilled water temperature set-point, dimming or turning off lights, decreasing duct static pressure set-points and restricting the supply fan operation \etc. 
In a large building, it is difficult to asses the effect of one control action on other sub-systems and on the building's overall power consumption because the building sub-systems are tightly coupled. 
Consider the case of the University of Pennsylvania's campus, which has over a hundred different buildings and centralized chiller plants. In order to perform campus wide DR, the facilities manager must account for several hundred thousand set-points and their impact on the different buildings. Therefore, it is extremely difficult for a human operator to accurately gauge the building's or a campus's response.
\item \textbf{Interpretability of modeling and control}: Predictive models for buildings, regardless how sophisticated, lose their effectiveness unless they can be interpreted by human experts and facilities managers in the field.
For \eg artificial neural networks (ANN) obscure physical control knobs and interactions and hence, are difficult to interpret by building facilities managers.
Therefore, the required solution must be transparent, human centric and highly interpretable.
\end{enumerate}

\begin{figure}
\centering
\includegraphics[width=0.85\columnwidth]{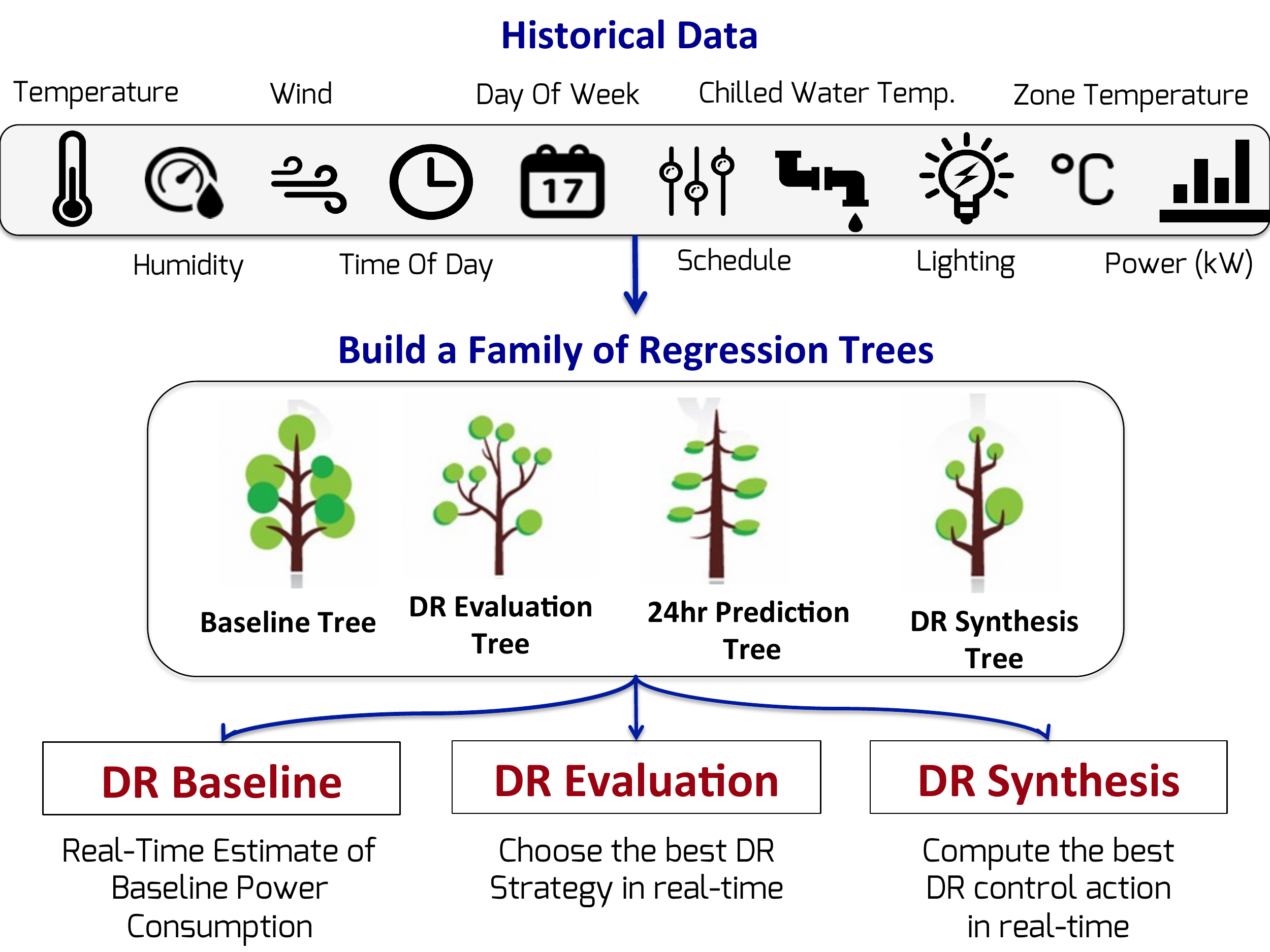}
\vspace{-3pt}
\caption{DR-Advisor Architecture}
\label{fig:overview}
\vspace{-16pt}
\end{figure}

The goal with data-driven methods for cyber-physical energy systems is to make the best of both worlds; i.e. simplicity of rule based approaches and the predictive capability of model based strategies, but without the expense of first principle or grey-box model development.

In this paper, we present a method called DR-Advisor (Demand Response-Advisor), which acts as a recommender system for the building's facilities manager and provides the power consumption prediction and control actions for meeting the required load curtailment and maximizing the economic reward.  
Using historical meter and weather data along with set-point and schedule information, DR-Advisor builds a family of interpretable regression trees to learn non-parametric data-driven models for predicting the power consumption of the building (Figure~\ref{fig:overview}).
DR-Advisor can be used for real-time demand response baseline prediction, strategy evaluation and control synthesis, without having to learn first principles based models of the building.

\subsection{Contributions}

This work has the following data-driven contributions:
\begin{enumerate}[leftmargin=0.5cm,topsep=1pt,itemsep=-1ex,partopsep=1ex,parsep=1ex]
\item \textbf{DR Baseline Prediction:} We demonstrate the benefit of using regression trees based approaches for estimating the demand response baseline power consumption. Using regression tree-based algorithms eliminates the cost of time and effort required to build and tune first principles based models of buildings for DR. DR-Advisor achieves a prediction accuracy of $92.8\%$ to $98.9\%$ for baseline estimates of eight buildings on the Penn campus.
\item \textbf{DR Strategy Evaluation:} We present an approach for building auto-regressive trees and apply it for demand response strategy evaluation. Our models takes into account the state of the building and weather forecasts to help choose the best DR strategy among several pre-determined strategies.
\item \textbf{DR Control Synthesis:}  We introduce a novel model based control with regression trees (mbCRT) algorithm to enable control with regression trees use it for real-time DR synthesis. Using the mbCRT algorithm, we can optimally trade off thermal comfort inside the building against the amount of load curtailment. While regression trees are a popular choice for prediction based models, this is the first time regression tree based algorithms have been used for controller synthesis with applications in demand response. Our synthesis algorithm outperforms rule based DR strategy by $17\%$ while maintaining bounds on thermal comfort inside the building.
\end{enumerate}

\subsection{Experimental validation and evaluation}
We evaluate the performance of DR-Advisor using a mix of real data from 8 buildings on the campus of the University of Pennsylvania, in Philadelphia USA and data-sets from a virtual building test-bed for the Department of Energy's (DoE) large commercial reference building.
We also compare the performance of DR-Advisor against other data-driven methods using a bench-marking data-set from AHRAE's great energy predictor shootout challenge.

This paper is organized as follows: Section~\ref{sec:problem} describes the challenges with demand response. 
In Section~\ref{sec:drtree}, we present how data-driven algorithms can be used for the problems associated with DR. 
Section~\ref{sec:drsyn}, presents a new algorithm to perform control with regression trees for synthesizing demand response strategies.
Section~\ref{sec:case} presents a comprehensive case study with DR-Advisor using data from several real buildings.
We conclude this paper in Section~\ref{sec:discussion} with a summary of our results and a discussion about future directions.

\section{Problem Definition}
\label{sec:problem}
\begin{figure}
  \centering
  \includegraphics[width=0.9\columnwidth]{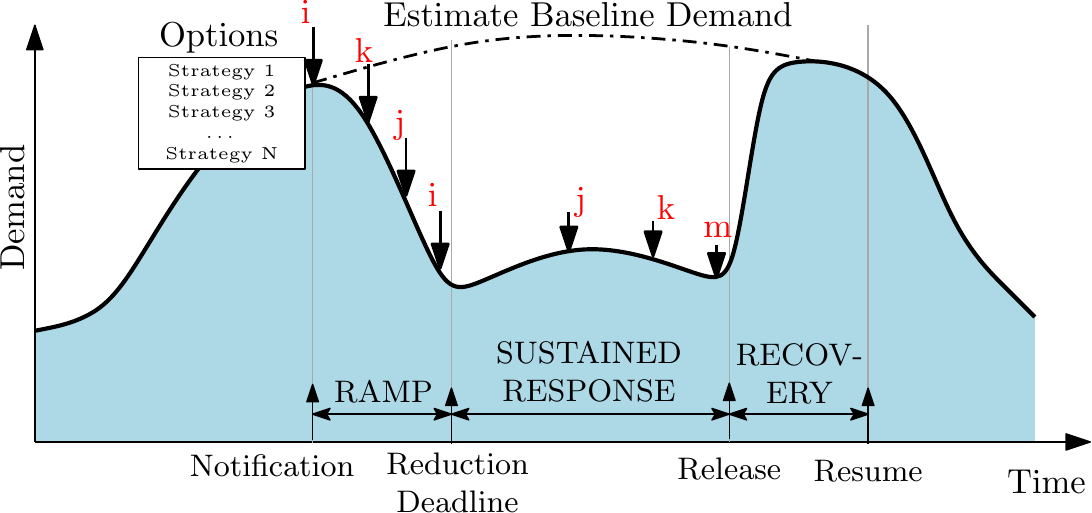}
  \vspace{-3pt}
  \caption{Example of a demand response timeline.}
  \label{fig:baseline}
   \vspace{-13pt}
\end{figure}

The timeline of a DR event is shown in Figure~\ref{fig:baseline}.
An \emph{event notification} is issued by the utility/CSP, at the notification time ($\sim$30mins).
The time by which the reduction must be achieved, is the \emph{reduction deadline}.
The main period during which the demand needs to be curtailed is the \emph{sustained response period} (1$\sim$6hrs).
The end of the response period is when the main curtailment is released. 
The normal operation is gradually resumed during the \emph{recovery period}.
The DR event ends at the end of the recovery period.

The key to answering the question of what actions to take to achieve a significant DR curtailment upon receiving a notification, lies in making accurate predictions about the power consumption response of the building. 
Specifically, it involves solving the three challenging problems of end-user demand response, which are described next.

\subsection{DR baseline prediction}
\label{sec:baselining}
The DR baseline is an estimate of the electricity that would have been consumed by a customer in the absence of a demand response event (as shown in In Fig.~\ref{fig:baseline}) 
The measurement and verification of the demand response baseline is the most critical component of any DR program since the amount of DR curtailment, and any associated financial reward can only be determined with respect to the baseline estimate.
The goal is to learn a predictive model $g()$ which relates the baseline power consumption estimate $\hat{Y_{base}}$ to the forecast of the weather conditions and building schedule for the duration of the DR-event \ie $\hat{Y_{base}} = g(\text{weather, schedule})$

\subsection{DR strategy evaluation}
\label{sec:eval}
Most DR today is manual and conducted using fixed rules and pre-determined curtailment strategies based on recommended guidelines, experience and best practices. 
During a DR event, the building's facilities manager must choose a single strategy among several pre-determined strategies to achieve the required power curtailment. 
Each strategy includes adjusting several control knobs such as temperature set-points, lighting levels and temporarily switching off equipment and plug loads to different levels across different time intervals. 

As only one strategy can be used at a time, the question then is, \emph{how to choose the DR strategy from a pre-determined set of strategies which leads to the largest load curtailment?}

Instead of predicting the baseline power consumption $\hat{Y_{base}}$, in this case we want the ability to predict the actual response of the building $\hat{Y_{kW}}$ due to any given strategy.
For example, in Fig.~\ref{fig:baseline}, there are $N$ different strategies available to choose from. 
DR-Advisor predicts the power consumption of the building due to each strategy and chooses the DR strategy ($ \in \{i,j, \cdots k \cdots N\}$) which leads to the largest load curtailment.
The resulting strategy could be a combination of switching between the available set of strategies.

\subsection{DR strategy synthesis}
\label{sec:synthesis}
Instead of choosing a DR strategy from a pre-determined set of strategies, a harder challenge is to 
synthesize new DR strategies and obtain optimal operating points for the different control variables.
We can cast this problem as an optimization over the set of control variables, $\mathbb{X}_c$, such that
\begin{equation}
\begin{aligned}
& \underset{\mathbb{X}_c}{\text{minimize}}
& & f(\hat{Y_{kW}}) \\
& \text{subject to}
& & \hat{Y_{kW}} = h(\mathbb{X}_c) \\
& & & \mathbb{X}_c \in \mathbb{X}_{safe}
\end{aligned}
\label{eq:linear_program}
\end{equation}
we want to minimize the predicted power response of the building $\hat{Y_{kW}}$, subject to a predictive model which relates the response to the control variables and subject to the constraints on the control variables.

Unlike rule-base DR, which does not account for building state and external factors, in DR synthesis the optimal control actions are derived based on the current state of the building, forecast of outside weather and electricity prices.


\section{Data-Driven Demand Response}
\label{sec:drtree}
Our goal is to find data-driven functional models that relates the value of the response variable, say power consumption, $\hat{Y_{kW}}$ with the values of the predictor variables or features $[X_1, X_2,\cdots, X_m]$ which can include weather data, set-point information and building schedules.
When the data has lots of features, as is the case in large buildings, which interact in complicated, nonlinear ways, assembling a single global model, such as linear or polynomial regression, can be difficult, and lead to poor response predictions.
An approach to non-linear regression is to partition the data space into smaller regions, where the interactions are more manageable. 
We then partition the partitions again; this is called recursive partitioning, until finally we get to chunks of the data space which are so tame that we can fit simple models to them. 
Regression trees is an example of an algorithm which belongs to the class of recursive partitioning algorithms. The seminal algorithm for learning regression trees is CART as described in~\cite{breiman1984classification}. 

Regression trees based approaches are our choice of data-driven models for DR-Advisor. The primary reason for this modeling choice is that regression trees are highly interpretable, by design.
Interpretability is a fundamental desirable quality in any predictive model.  
Complex predictive models like neural-networks , support vector regression \etc go through a long calculation routine and involve too many factors. 
It is not easy for a human engineer to judge if the operation/decision is correct or not or how it was generated in the first place. 
Building operators are used to operating a system with fixed logic and rules. 
They tend to prefer models that are more transparent, where it is clear exactly which factors were used to make a particular prediction.
At each node in a regression tree a simple, if this then that, human readable, plain text rule is applied to generate a prediction at the leafs, which anyone can easily understand and interpret.
Making machine learning algorithms more interpretable is an active area of research~\cite{giraud1998beyond}, one that is essential for incorporating human centric models in cyber-physical energy systems.

\subsection{Data-Description}

In order to build regression trees which can predict the power consumption of the building, we need to train on time-stamped historical data. As shown in Fig.~\ref{fig:overview}, the data that we use can be divided into three different categories as described below:
\begin{enumerate}[leftmargin=0.5cm,topsep=1pt,itemsep=-1ex,partopsep=1ex,parsep=1ex]
\item \textbf{Weather Data:} It includes measurements of the outside dry-bulb and wet-bulb air temperature, relative humidity, wind characteristics and solar irradiation at the building site.
\item \textbf{Schedule data:} We create \emph{proxy} variables which correlate with repeated patterns of electricity consumption \eg due to occupancy or equipment schedules. \emph{Day of Week} is a categorical predictor which takes values from $1-7$ depending on the day of the week. This variable can capture any power consumption patterns which occur on specific days of the week. For instance, there could a big auditorium in an office building which is only used on certain days. Likewise, \emph{Time of Day} is quite an important predictor of powe consumption as it can adequately capture daily patterns of occupancy, lighting and appliance use without directly measuring any one of them. Besides using proxy schedule predictors, actual building equipment schedules can also be used as training data for building the trees.
\item \textbf{Building data:} The state of the building is required for DR strategy evaluation and synthesis. This includes (i) Chilled Water Supply Temperature (ii) Hot Water Supply Temperature (iii) Zone Air Temperature (iv) Supply Air Temperature (v) Lighting levels.
\end{enumerate}

\subsection{Data-Driven DR Baseline}

DR-Advisor uses a mix of several algorithms to learn a reliable baseline prediction model. For each algorithm, we train the model on historical power consumption data and then validate the predictive capability of the model against a test data-set which the model has never seen before.
In addition to building a single regression tree, we also learn cross-validated regression trees, boosted regression trees (BRT) and random forests (RF). The ensemble methods like BRT and RF help in reducing any over-fitting over the training data. They achieve this by combining the predictions of several base estimators built with a given learning algorithm in order to improve generalizability and robustness over a single estimator.
For a more comprehensive review of random forests we refer the reader to~\cite{breiman2001random}.
A boosted regression tree (BRT) model is an additive regression model in which individual terms are simple trees, fitted in a forward, stage-wise fashion~\cite{elith2008working}.

\subsection{Data-Driven DR Evaluation}
\label{sec:autort}

The regression tree models for DR evaluation are similar to the models used for DR baseline estimation except for two key differences:
First, instead of only using weather and proxy variables as the training features, in DR evaluation, we also train on set-point schedules and data from the building itself to capture the influence of the state of the building on its power consumption; and 
Second, in order to predict the power consumption of the building for the entire length of the DR event, we use the notion of auto-regressive trees. An auto-regressive tree model is a regular regression tree except that the lagged values of the response variable are also predictor variables for the regression tree \ie the tree structure is learned to approximate the following function:
\begin{equation}
\hat{Y_{kW}(t)} = f([X_1, X_2,\cdots, X_m,Y_{kW}(t-1),\cdots,Y_{kW}(t-\delta)])
\end{equation}
where the predicted power consumption response $\hat{Y_{kW}}$ at time $t$, depends on previous values of the response itself $[Y_{kW}(t-1),\cdots,Y_{kW}(t-\delta)]$ and $\delta$ is the order of the auto-regression.
This allows us to make finite horizon predictions of power consumption for the building.
At the beginning of the DR event we use the auto-regressive tree for predicting the response of the building due to each rule-based strategy and choose the one which performs the best over the predicted horizon. The prediction and strategy evaluation is re-computed periodically throughout the event.

\section{Data-Driven Control Synthesis}
\label{sec:drsyn}
The data-driven methods described so far use the forecast of features to obtain building power consumption predictions  for DR baseline and DR strategy evaluation.
In this section, we extend the theory of regression trees to solve the demand response synthesis problem described earlier in Section~\ref{sec:synthesis}. This is our primary contribution. 
  \begin{figure}
  \centering
  \includegraphics[width=0.5\columnwidth]{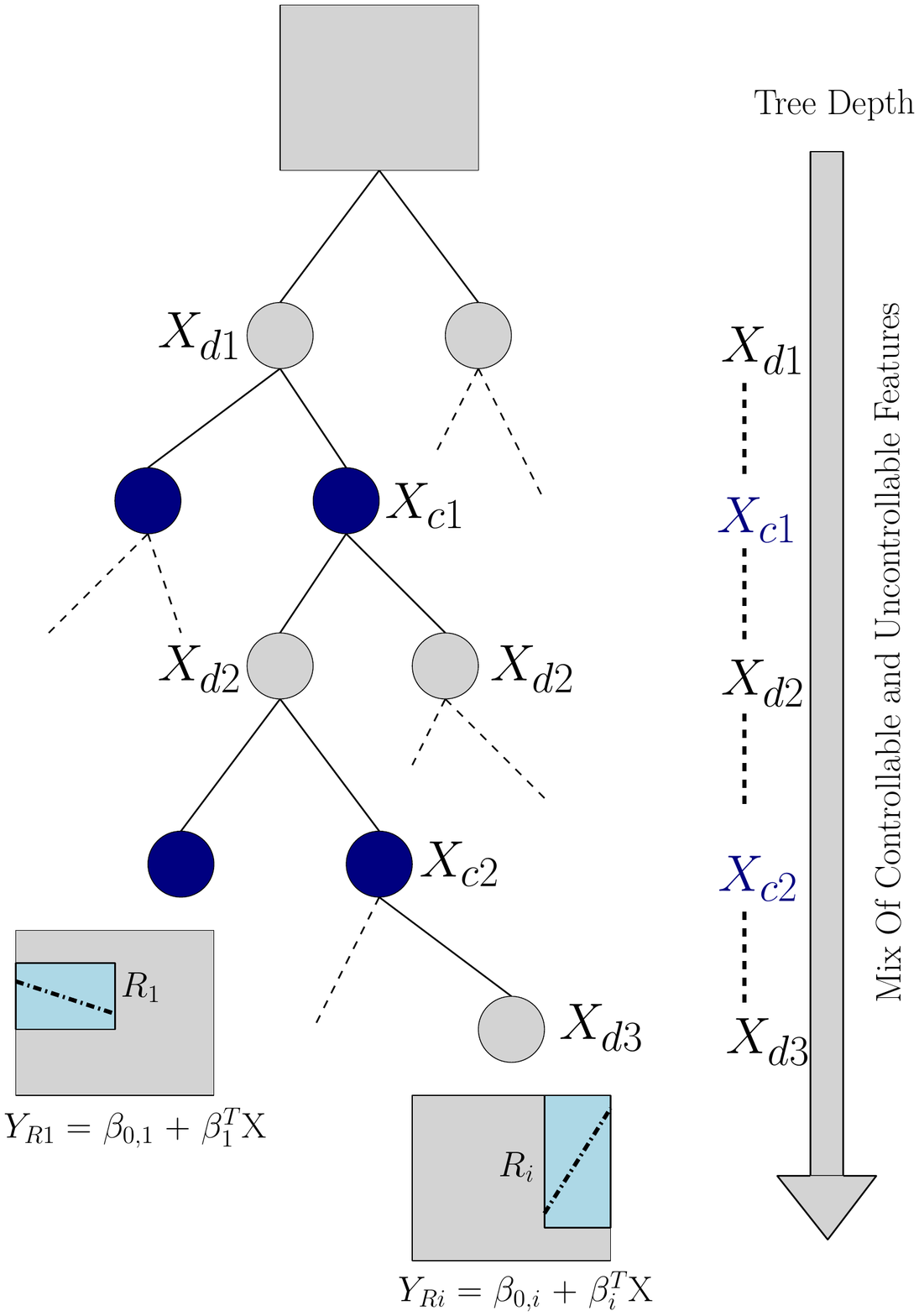}
  \caption{Example of a regression tree with linear regression model in leaves. Not suitable for control due to the mixed order of the controllable $X_c$ (solid blue) and uncontrollable $X_d$ features.}
  \label{fig:mixed_order}
  \vspace{-10pt}
\end{figure}

Recall that the objective of learning a regression tree is to learn a model $f$ for predicting the response $Y$ with the values of the predictor variables or features $X_1, X_2,\cdots, X_m$; \ie $Y=f([X_1, X_2,\cdots, X_m])$
Given a forecast of the features $\hat{X_1}, \hat{X_2},\cdots, \hat{X_m}$ we can predict the response $\hat{Y}$. 
Now consider the case where a subset, $\mathbb{X}_c \subset \mathbb{X}$ of the set of features/variables $\mathbb{X}$'s are manipulated variables \ie we can change their values in order to drive the response $(\hat{Y})$ towards a certain value. 
In the case of buildings, the set of  variables can be separated into disturbances (or non-manipulated) variables like outside air temperature, humidity, wind etc. while the controllable (or manipulated) variables would be the temperature and lighting set-points within the building.
Our goal is to modify the regression trees and make them suitable for synthesizing the optimal values of the control variables in real-time.

\subsection{Model-based control with regression trees}
\label{sec:mbcrt}

The key idea in enabling control synthesis for regression trees is in the separation of features/variables into manipulated and non-manipulated features. 
Let $\mathbb{X}_c \subset \mathbb{X}$ denote the set of manipulated variables and $\mathbb{X}_d \subset \mathbb{X}$ denote the set of disturbances/ non-manipulated variables such that $\mathbb{X}_c \cup \mathbb{X}_d \equiv \mathbb{X}$.
Using this separation of variables we build upon the idea of simple model based regression trees~\cite{quinlan1992learning, friedman1991multivariate} to \emph{model based control with regression trees (mbCRT)}. 

Figure~\ref{fig:mixed_order} shows an example of how manipulated and non-manipulated features can get distributed at different depths of model based regression tree which uses the a linear regression function in the leaves of the tree:
\begin{equation}
\hat{Y_{Ri}} = \beta_{0,i} + \beta^T_i \mathbb{X}
\label{eq:linear_regression_leaf}
\end{equation}
Where $\hat{Y_{Ri}}$ is the predicted response in region $R_i$ of the tree using all the features $\mathbb{X}$. 
 In such a tree the prediction can only be obtained if the values of all the features $X$'s is known, including the values of the control variables $X_{ci}$'s. 
Since the manipulated and non-manipulated variables appear in a mixed order in the tree depth, we cannot use this tree for control synthesis.
This is because the value of the control variables $X_{ci}$'s is unknown, one cannot navigate to any single region using the forecasts of disturbances alone. 
\begin{figure}
  \centering
  \includegraphics[width=0.8\columnwidth]{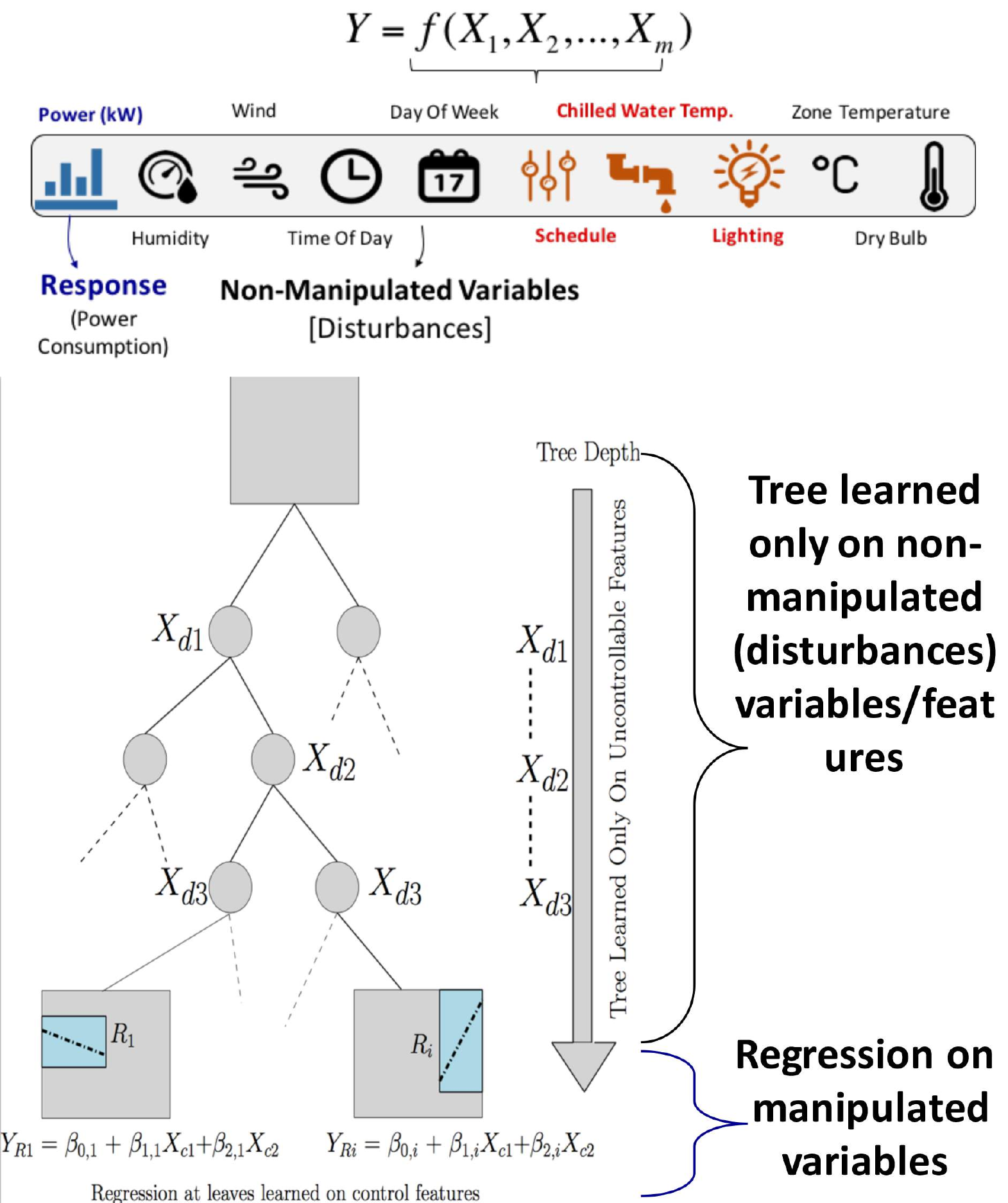}
  \caption{Example of a tree structure obtained using the mbCRT algorithm. The separation of variables allows using the linear model in the leaf to use only control variables.}
  \label{fig:algo1}
   \vspace{-5pt}
\end{figure}

The mbCRT algorithm avoids this problem using a simple but clever idea. We still partition the entire data space into regions using CART algorithm, but the top part of the regression tree is learned only on the non-manipulated features $\mathbb{X}_d$ or disturbances as opposed to all the features $\mathbb{X}$ (Figure~\ref{fig:algo1})
In every region at the leaves of the ``disturbance'' tree a linear model is fit but only on the control variables $\mathbb{X}_c$:
\begin{equation}
Y_{Ri} = \beta_{0,i} + \beta^T_i \mathbb{X}_c
\label{eq:control_leaf}
\end{equation}
Separation of variables allows us to use the forecast of the disturbances $\hat{\mathbb{X}_d}$ to navigate to the appropriate region $R_i$ and use the linear regression model ($Y_{Ri} = \beta_{0,i} + \beta^T_i \mathbb{X}_c$) with only the control/manipulated features in it as the valid prediction model for that time-step.

\small
\begin{algorithm}
\caption{mbCRT: Model Based Control With Regression Trees}\label{alg:mbcrt}
\begin{algorithmic}[1]
\State \textsc{Design Time}
\Procedure{Model Training}{}
\State \textit{Separation of Variables}
\State \textit{Set} $\mathbb{X}_c$ $\gets$ non-manipulated features
\State \textit{Set} $\mathbb{X}_d$ $\gets$ manipulated features
\State Build  the power prediction tree $T_{kW}$ with $\mathbb{X}_d$
\ForAll{Regions $R_i$ at the leaves of $T_{kW}$}
\State Fit linear model $\hat{kW}_{Ri} = \beta_{0,i} + \beta^T_i \mathbb{X}_c$
\State Build $q$ temperature trees $T1,T2 \cdots Tq$ with $\mathbb{X}_d$
\EndFor
\ForAll{Regions $R_i$ at the leaves of $Ti$}
\State Fit linear model $\hat{Ti} = \beta_{0,i} + \beta^T_i \mathbb{X}_c$
\EndFor
\EndProcedure
\State \textsc{Run Time}
\Procedure{Control Synthesis}{}
\State At time t obtain forecast $\hat{\mathbb{X}_d}(t+1)$ of disturbances $\hat{X_{d1}}(t+1), \hat{X_{d2}}(t+1),\cdots$
\State Using $\hat{\mathbb{X}_d}(t+1)$ determine the leaf and region $R_{rt}$ for each tree.
\State Obtain the linear model at the leaf of each tree.
\State Solve optimization in Eq\ref{eq:synth_program} for optimal control action $\mathbb{X}^*_c(t)$
\EndProcedure
\end{algorithmic}
\end{algorithm}
\normalsize

\begin{figure}
\centering
\includegraphics[width=\columnwidth]{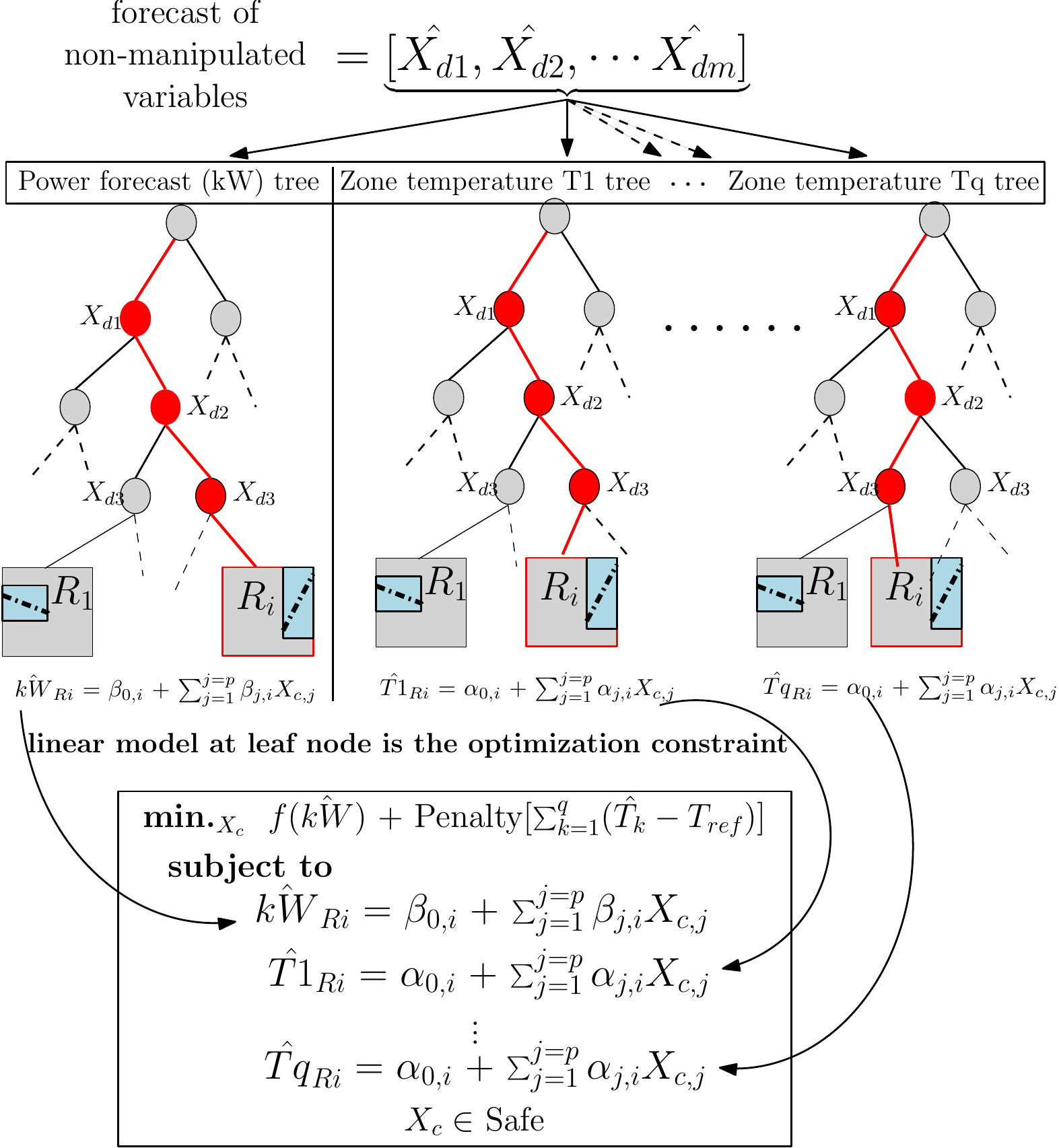}
\caption{DR synthesis with thermal comfort constraints. Each tree is responsible for contributing one constraint tot the demand response optimization.}
\label{fig:dropt}
\end{figure}

\subsection{DR synthesis optimization}
In the case of DR synthesis for buildings, the response variable is power consumption, the objective function can denote the financial reward of minimizing the power consumption during the DR event. However, the curtailment must not result in high levels of discomfort for the building occupants. In order to account for thermal comfort, in addition to learning the tree for power consumption forecast, we can also learn different trees to predict the temperature of different zones in the building. As shown in Figure~\ref{fig:dropt} and Algorithm~\ref{alg:mbcrt}, at each time-step during the DR event, a forecast of the non manipulated variables is used by each tree, to navigate to the appropriate leaf node. For the power forecast tree, the linear model at the leaf node relates the predicted power consumption of the building to the manipulated/control variables \ie $\hat{kW} = \beta_{0,i} + \beta^T_i \mathbb{X}_c$.

Similarly, for each zone $1,2,\cdots q$, a tree is built whose response variable is the zone temperature $Ti$. The linear model at the leaf node of each of the zone temperature tree relates the predicted zone temperature to the manipulated variables $\hat{Ti} = \alpha_{0,j} + \beta^T_j \mathbb{X}_c$.
Therefore, at every time-step, based on the forecast of the non-manipulated variables, we obtain $q+1$ linear models between the power consumption and $q$ zone temperatures and the manipulated variables. We can then solve the following DR synthesis optimization problem to obtain the values of the manipulated variables $\mathbb{X}_c$:
\begin{center}
\begin{equation}
\begin{aligned}
\underset{\mathbb{X}_c}{\text{minimize}}
 & f(\hat{kW})+\text{Penalty}[\sum_{k=1}^q(\hat{T_k}-T_{ref})] \\
\text{subject to} \\
& \hat{kW} = \beta_{0,i} + \beta^T_i \mathbb{X}_c \\
& \hat{T1} = \alpha_{0,1} + \beta^T_1 \mathbb{X}_c \\
& \cdots \\
& \hat{Td} = \alpha_{0,q} + \beta^T_q \mathbb{X}_c \\
& \mathbb{X}_c \in \mathbb{X}_{safe}
\end{aligned}
\label{eq:synth_program}
\end{equation}
\end{center}
\begin{figure}
  \begin{subfigure}
    \centering
  \includegraphics[width=\columnwidth]{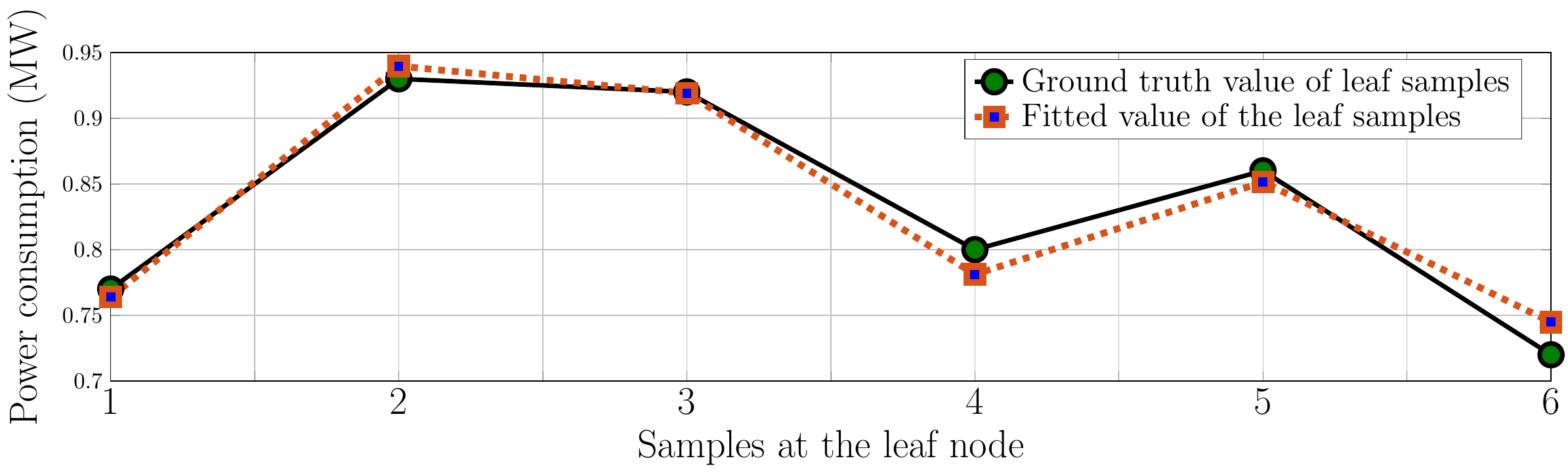}
  \end{subfigure}  
  \begin{subfigure}
    \centering
  \includegraphics[width=\columnwidth]{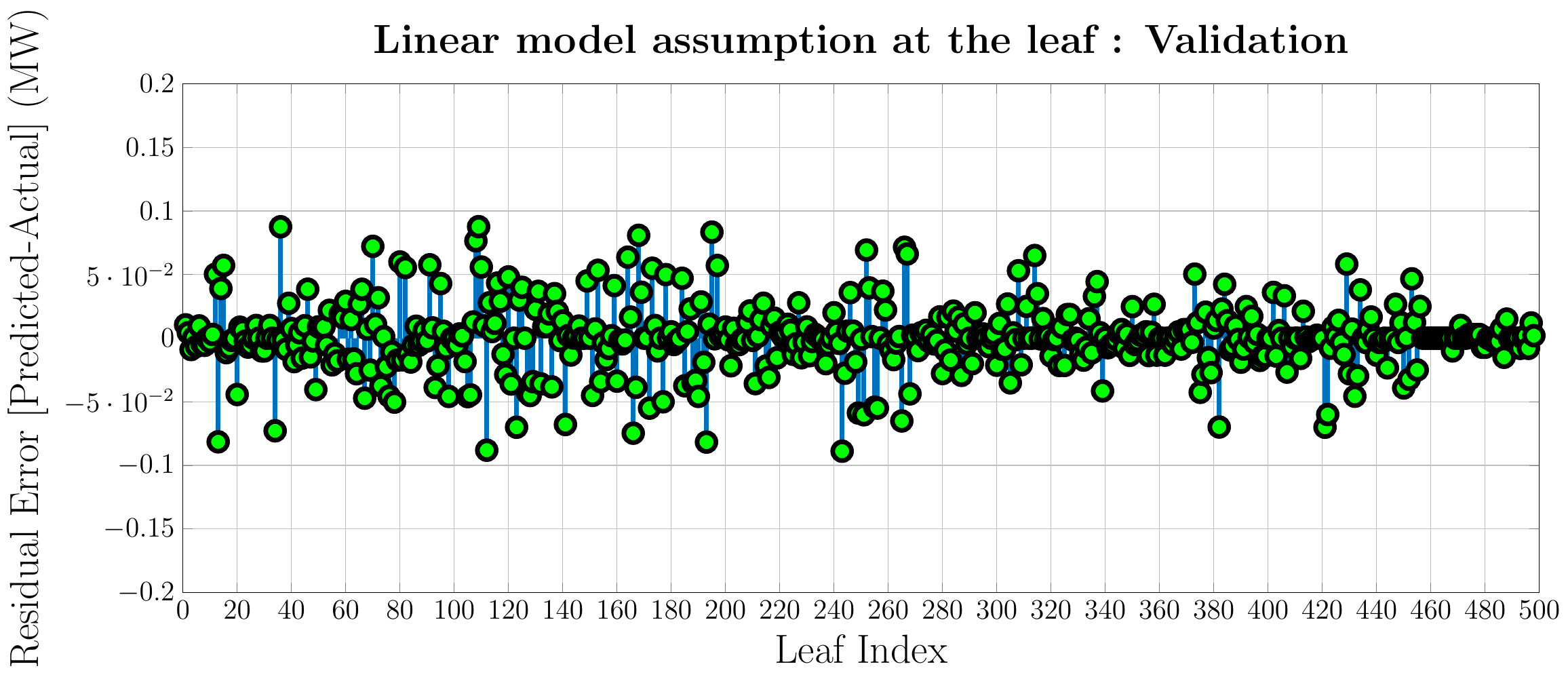}
  \end{subfigure}
  \caption{Linear model assumption at the leaves}
  \label{fig:leaf}
   \vspace{-5pt}
\end{figure}

The linear model between the response variable $Y_{Ri}$ and the control features $\mathbb{X}_c$ is assumed for computational simplicity. Other models could also be used at the leaves as long as they adhere to the separation of variables principle. Figure~\ref{fig:leaf} shows that the linear model assumption in the leaves of the tree is a valid assumption.

The intuition behind the mbCRT Algorithm~\ref{alg:mbcrt} is that at run time $t$, we use the forecast $\hat{\mathbb{X}_d}(t+1)$ of the disturbance features to determine the region of the \textit{uncontrollable} tree and hence, the linear model to be used for the control.
We then solve the simple linear program corresponding to that region to obtain the optimal values of the control variables.

The mbCRT algorithm is the first ever algorithm which allows the use of regression trees for control synthesis.

\section{Case Study}
\label{sec:case}
\begin{figure}
\centering
\includegraphics[width=\columnwidth]{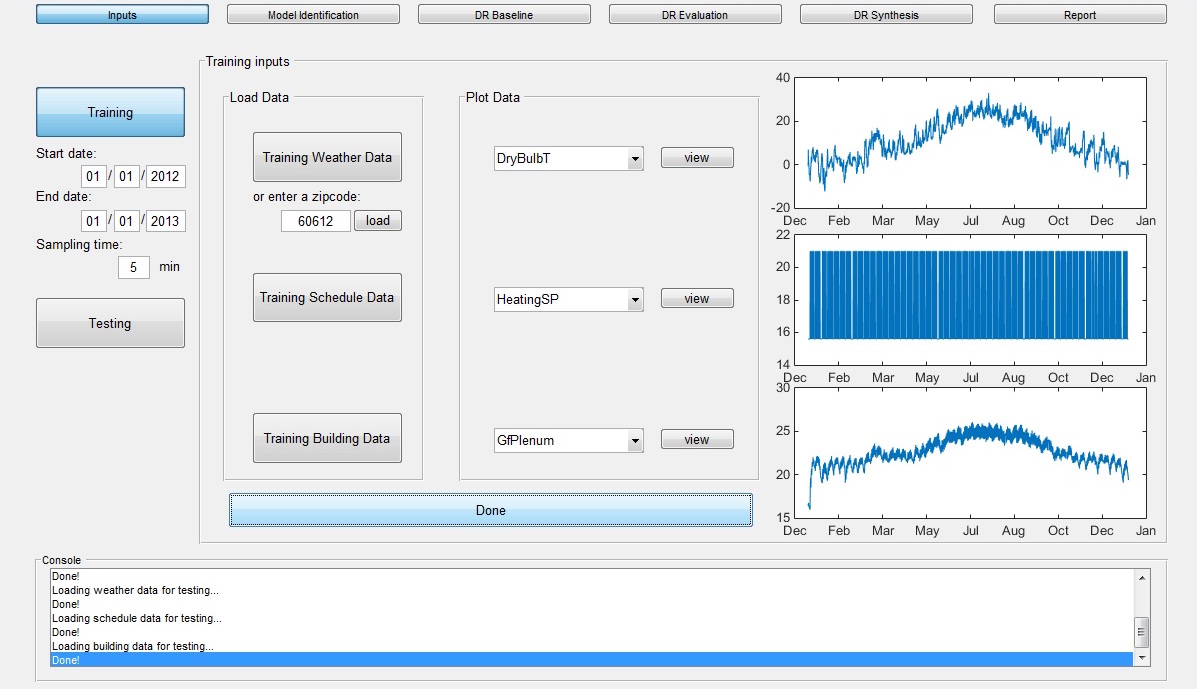}
  \caption{DR-Advisor toolbox for power, price, weather and schedule data capture, baseline prediction, DR policy evaluation and DR synthesis}
  \label{gui}
\end{figure}

DR-Advisor (Figure~\ref{gui}) has been developed into a MATLAB toolbox available at \url{http://mlab.seas.upenn.edu/dr-advisor/}.
In this section, we present a comprehensive case study to show how DR-Advisor can be used to address all the aforementioned demand response challenges (Section~\ref{sec:problem}) and we compare the performance of our tool with other data-driven methods. 

\subsection{Building description}
\begin{figure}[b]
\centering
\includegraphics[width=0.9\columnwidth]{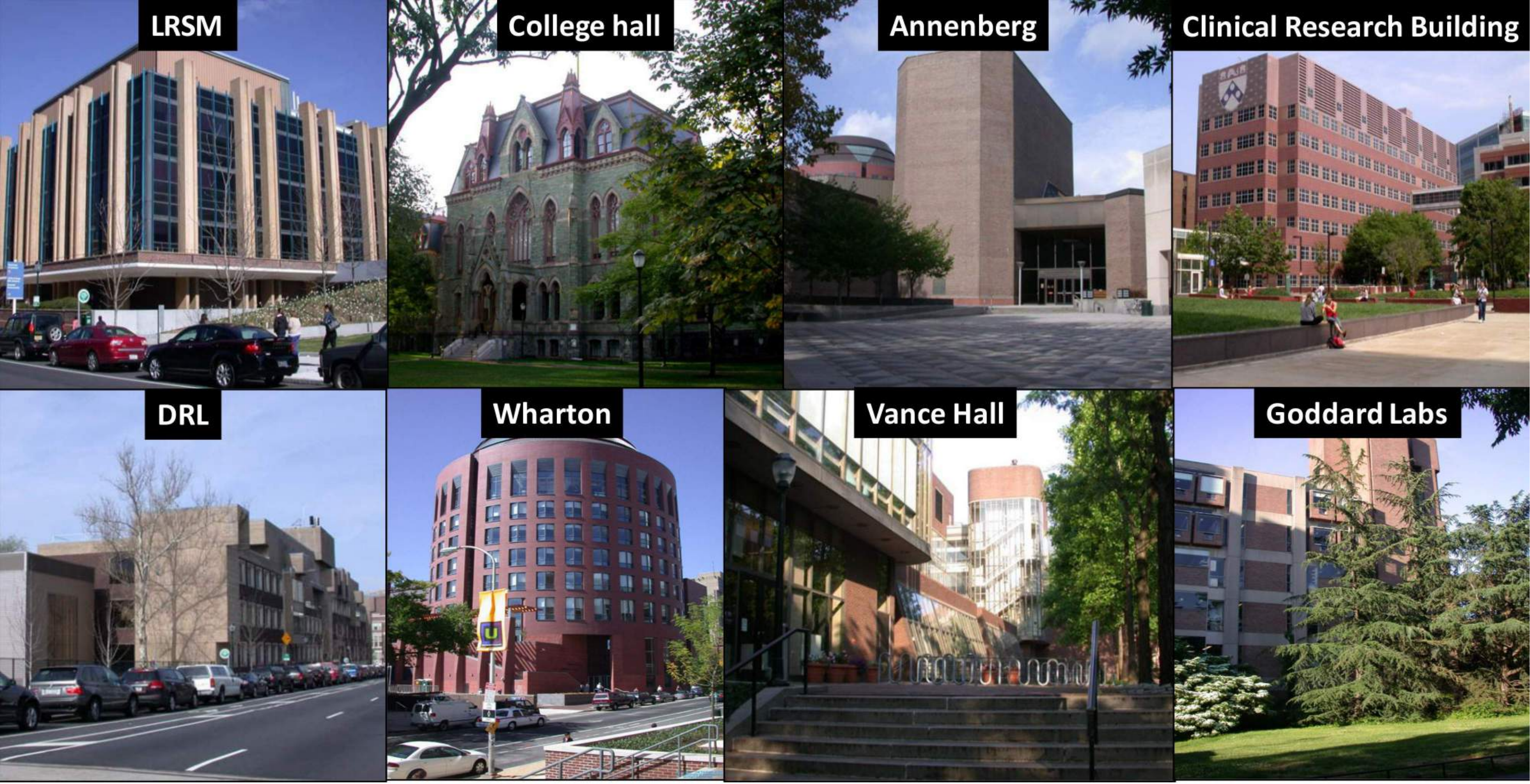}
\caption{8 different buildings on Penn campus were modeled with DR-Advisor}
\label{fig:penn}
\end{figure}

We use historical weather and power consumption data from 8 buildings on the Penn campus (Figure~\ref{fig:penn}). These buildings are a mix of scientific research labs, administrative buildings, office buildings with lecture halls and bio-medical research facilities. The total floor area of the eight buildings is over $1.2$ million square feet spanned across. The size of each building is shown in Table~\ref{tab:penn}.

We also use the DoE Commercial Reference Building (DoE CRB) simulated in EnergyPlus~\cite{referencebuilding} as the virtual test-bed building.
This is a large 12 story office building consisting of $73$ zones with a total area of $500,000$ sq ft. 
There are ~$2,397$ people in the building during peak occupancy. 
During peak load conditions the building can consume up to $1.6$ MW of power. 
For the simulation of the DoE CRB building we use actual meteorological year data from Chicago for the years $2012$ and $2013$. 
On July 17, 2013, there was a DR event on the PJM ISO grid from 15:00 to 16:00 hrs. We simulated the DR event for the same interval for the virtual test-bed building.

\subsection{Model Validation}
For each of the Penn buildings, multiple regression trees were trained on weather and power consumption data from August 2013 to  December 2014. 
Only the weather forecasts and proxy variables were used to train the models.
We then use the DR-Advisor to predict the power consumption in the test period \ie for several months in 2015. 
The predictions are obtained for each hour, making it equivalent to baseline power consumption estimate. 
The predictions on the test-set are compared to the actual power consumption of the building during the test-set period. 
\begin{figure}
\centering
\includegraphics[height=3cm, width=\columnwidth]{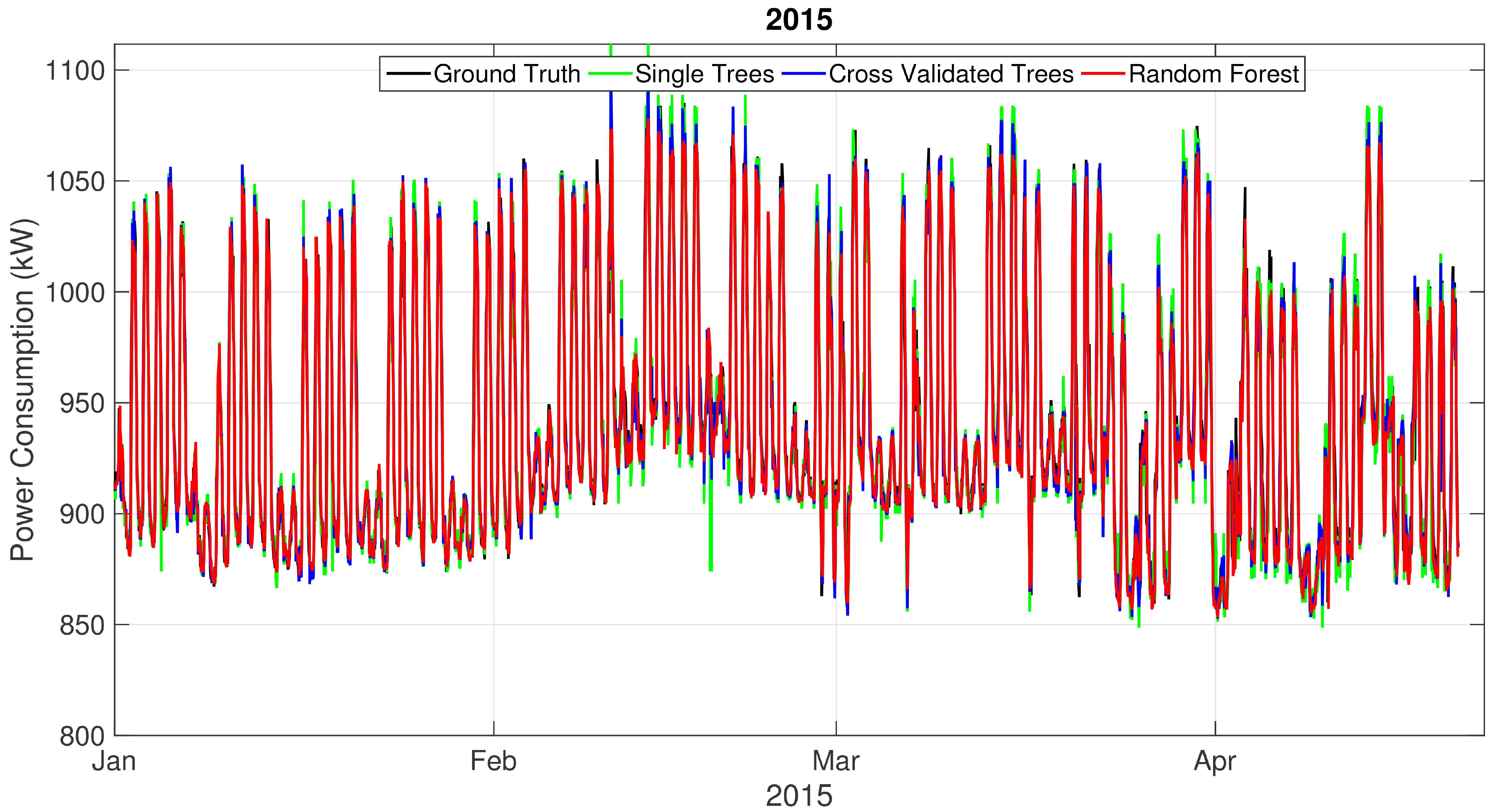}
\caption{Model validation for the clinical research building at Penn.}
\label{fig:clinical}
\end{figure}
One such comparison for the clinical reference building is shown in Figure~\ref{fig:clinical}. The following algorithms were evaluated: single regression tree, k-fold cross validated (CV) trees, boosted regression trees (BRT) and random forests (RF).
Our chosen metric of prediction accuracy is the one minus the normalized root mean square error (NRMSE). NRMSE is the RMSE divided by the mean of the data. The accuracy of the model of all the eight buildings is summarized in Table~\ref{tab:penn}. We notice that DR-Advisor performs quite well and the accuracy of the baseline model is between $92.8\%$ to $98.9\%$ for all the buildings.
\begin{table}
\caption{Model validation with Penn data}
\vspace{3pt}
\resizebox{\columnwidth}{!}{%
    \begin{tabular}{|l|c|c|c|}
    \hline
    \textbf{Building Name }             & \textbf{Total Area (sq-ft)} & \textbf{Floors} & \textbf{Accuracy (\%)} \\ \hline
    LRSM                       & 92,507             & 6      & 94.52       \\
    College Hall               & 110,266            & 6      & 96.40       \\
    Annenberg Center           & 107,200            & 5      & 93.75       \\
    Clinical Research Building & 204,211            & 8      & 98.91       \\
    David Rittenhouse Labs     & 243,484            & 6      & 97.91       \\
    Huntsman Hall              & 320,000            & 9      & 95.03       \\
    Vance Hall                 & 106,506            & 7      & 92.83       \\
    Goddard Labs               & 44,127             & 10     & 95.07       \\ \hline
    \end{tabular}
    }
 \label{tab:penn}   
\end{table}

\subsection{Energy Prediction Benchmarking}
\label{sec:ashrae}
We compare the performance of DR-Advisor with other data-driven method using a bench-marking data-set from the American Society of Heating, Refrigeration and Air Conditioning Engineers (ASHRAE's) Great Energy Predictor Shootout Challenge~\cite{kreider1994predicting}. 
The goal of the ASHRAE challenge  was to explore and evaluate data-driven models that may not have such a strong physical basis, yet that perform well at prediction.
The competition attracted $\sim150$ entrants, who attempted to predict the unseen power loads from weather and solar radiation data using a variety of approaches.
In addition to predicting the hourly whole building electricity consumption, WBE (kW), both the hourly chilled water, CHW (millions of Btu/hr) and hot water consumption, HW (millions of Btu/hr) of the building was also required to be a prediction output. Four months of training data with the following features was provided: 
\begin{inparaenum}(a)
\item Outside temperature ($^\circ$F)
\item Wind speed (mph)
\item Humidity ratio (water/dry air)
\item Solar flux (W/$m^2$)
\end{inparaenum}
In addition to these training features, we added three proxy variables of our own: hour of day, IsWeekend and IsHoliday to account for correlation of the building outputs with schedule. 

\begin{table}
\centering
\caption{ASHRAE Energy Prediction Competition Results}
\vspace{3pt}
\resizebox{\columnwidth}{!}{%
    \begin{tabular}{|c|c|c|c|c|}
    \hline
    ASHRAE Team ID & WBE CV & CHW CV & HW CV & Average CV \\ \hline
    9              & 10.36  & 13.02  & 15.24 & 12.87      \\ \hline
    \textbf{DR-Advisor} & \textbf{11.72}  & \textbf{14.88}  & \textbf{28.13} & \textbf{18.24}     \\ \hline
    6              & 11.78  & 12.97  & 30.63 & 18.46      \\ \hline
    3              & 12.79  & 12.78  & 30.98 & 18.85      \\ \hline
    2              & 11.89  & 13.69  & 31.65 & 19.08      \\ \hline
    7              & 13.81  & 13.63  & 30.57 & 19.34      \\ \hline
    \end{tabular}
    }
    \label{tab:results}
\end{table}

Finally, we use different ensemble methods within DR-Advisor to learn models for predicting the three different building attributes. 
In the actual competition, the winners were selected based on the accuracy of all predictions as measured by the normalized root mean square error, also referred to as the coefficient of variation statistic CV. 
The smaller the value of CV, the better the prediction accuracy.
ASHRAE released the results of the competition for the top 19 entries which they received. 
In Table~\ref{tab:results}, we list the performance of the top 5 winners of the competition and compare our results with them.
It can be seen from table~\ref{tab:results}, that the random forest implementation in the DR-Advisor tool ranks $2^{nd}$ in terms of WBE CV and the overall average CV. The winner of the competition was an entry from David Mackay~\cite{mackay1994bayesian} which used a particular form of bayesian modeling using neural networks.

The result we obtain clearly demonstrates that the regression tree based approach within DR-Advisor can generate predictive performance that is comparable with the ASHRAE competition winners. Furthermore, since regression trees are much more interpretable than neural networks, their use for building electricity prediction is, indeed, very promising.

\subsection{DR-Evaluation}
\label{sec:case_eval}

\begin{figure}
\centering
\includegraphics[height=5cm, width=\columnwidth]{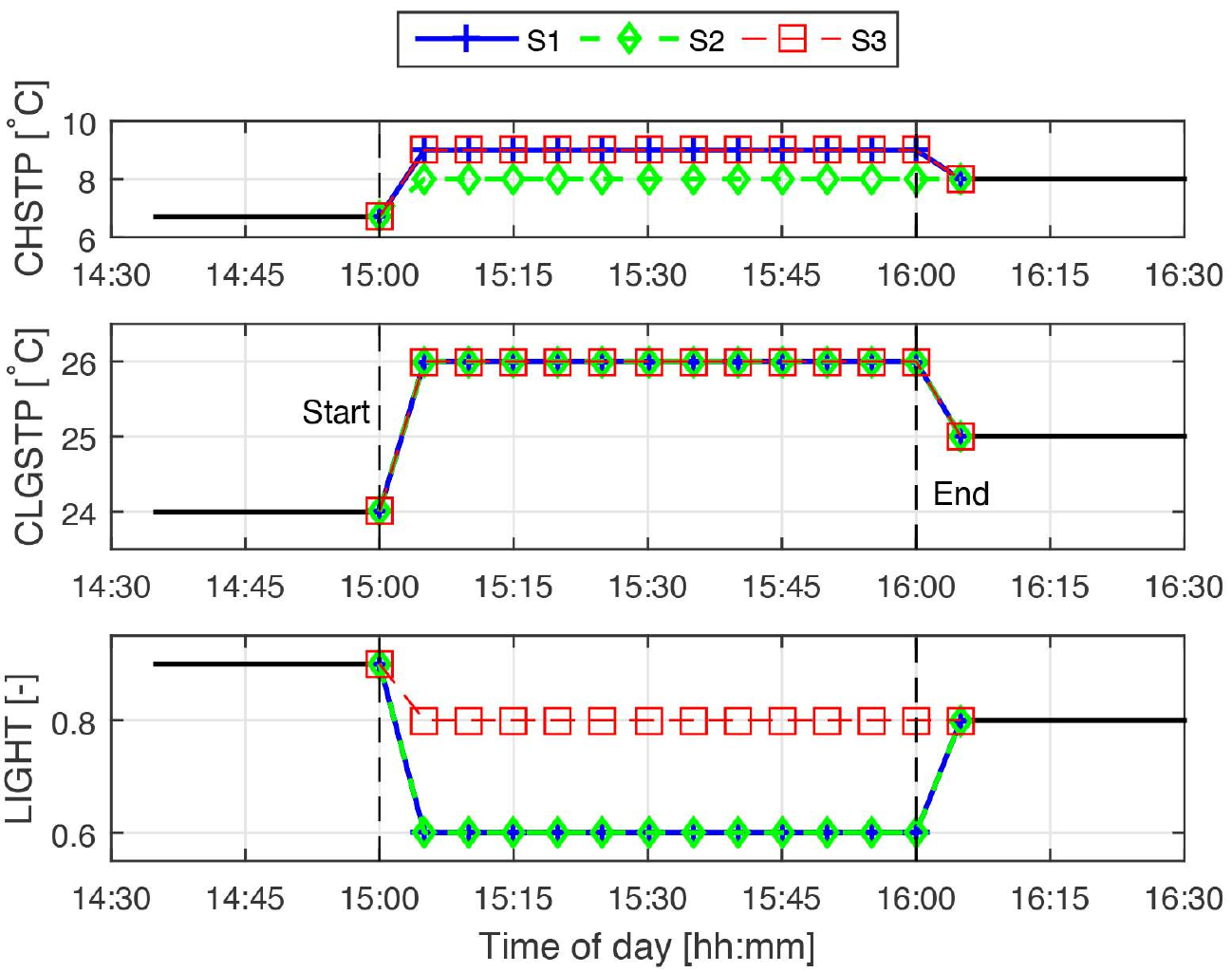}
\caption{Rule-based strategies used in DR Evaluation. CHSTP denotes Chiller set point and CLGSTP denotes Zone Cooling temperature set point. }
\label{fig:case_eval_control}
\vspace{-10pt}
\end{figure}

We test the performance of 3 different rule based strategies shown in Fig. \ref{fig:case_eval_control}. 
Each strategy determines the set point schedules for chiller water, zone temperature and lighting during the DR event. 
These strategies were derived on the basis of automated DR guidelines provided by Siemens~\cite{siemensdr}.
Chiller water set point is same in Strategy 1 (S1) and Strategy 3 (S3), higher than that in Strategy 2 (S2). Lighting level in S3 is higher than in S1 and S2.

We use auto-regressive trees (Section~\ref{sec:autort}) with order, $\delta = 6$ to predict the power consumption for the entire duration (1 hour) at the start of DR Event. In addition to learning the tree for power consumption, additional auto-regressive trees are also built for predicting the zone temperatures of the building.
At every time step, first the zone temperatures are predicted using the trees for temperature prediction. 
Then the power tree uses this temperature forecast along with lagged power consumption values to predict the power consumption recursively until the end of the prediction horizon.

\begin{figure}
\centering
\includegraphics[height=4cm, width=\columnwidth]{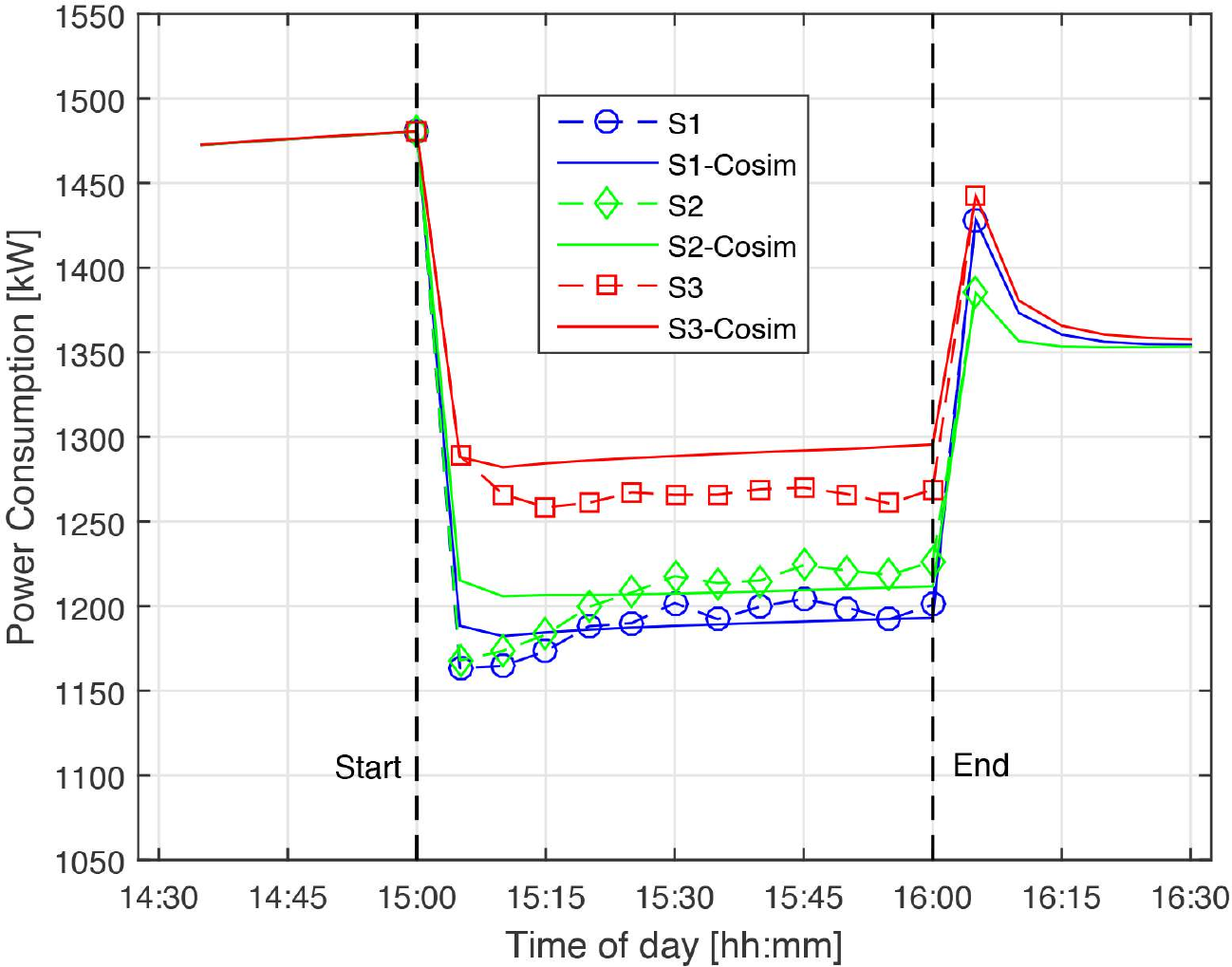}
\caption{Prediction of power consumption for 3 strategies. DR Evaluation shows that Strategy 1 (S1) leads to maximum power curtailment.}
\label{fig:case_eval_power}
\vspace{-10pt}
\end{figure}

Fig. \ref{fig:case_eval_power} shows the power consumption prediction using the auto-regressive trees and the ground truth obtained by simulation of the DoE CRB virtual test-bed for each rule-based strategy. 
Based on the predicted response, in this case DR-Advisor chooses to deploy the strategy S1, since it leads to the least amount of electricity consumption. The predicted response due to the chosen strategy aligns well with the ground truth power consumption of the building due to the same strategy, showing that DR strategy evaluation prediction of DR-Advisor is reliable and can be used to choose the best rule-based strategy from a set of pre-determined rule-based DR strategies.

\subsection{DR-Synthesis}
\label{sec:case_syn}

 \begin{figure}
\centering
\includegraphics[width=\columnwidth]{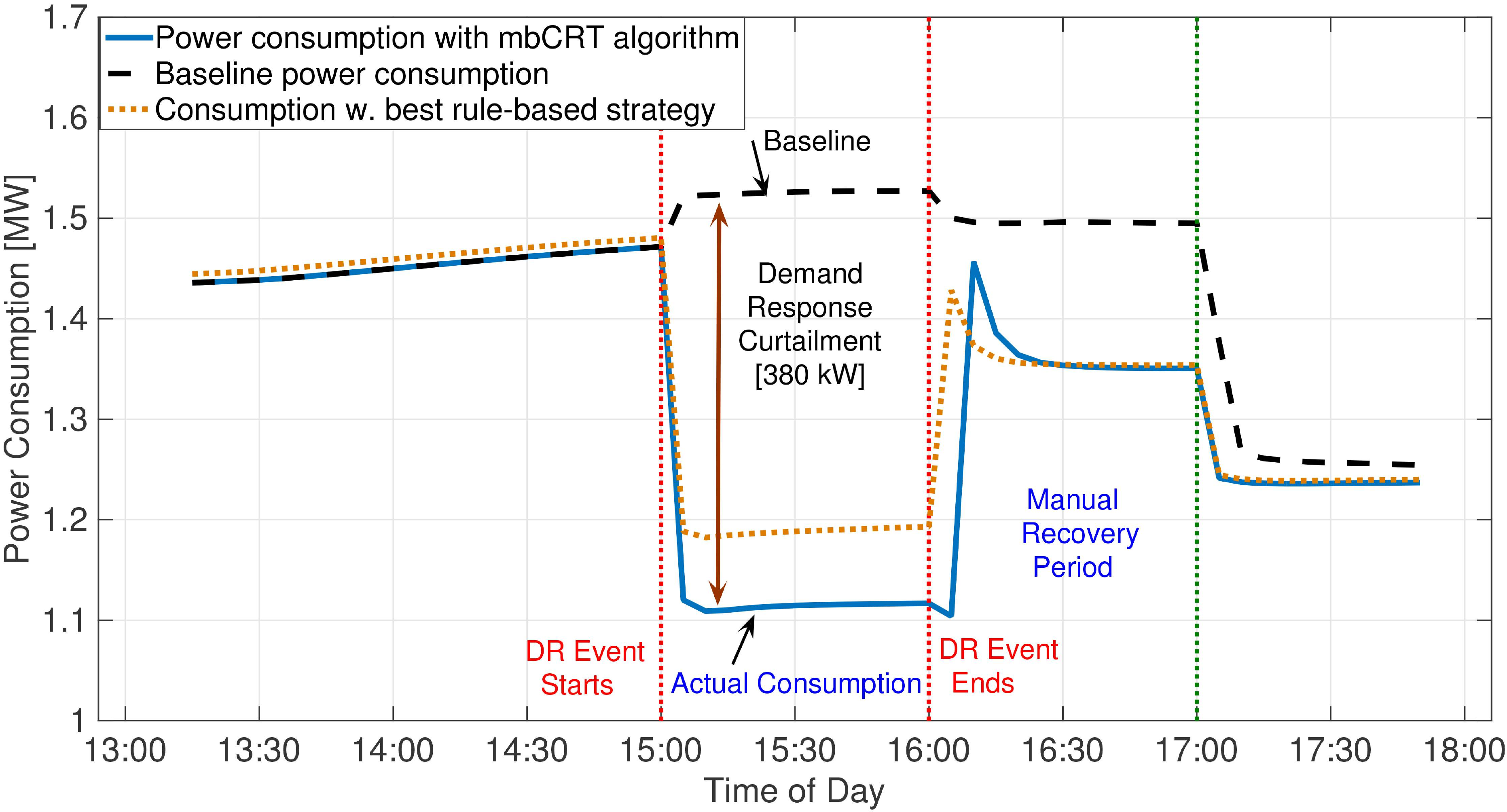}
\caption{DR synthesis using the mbCRT algorithm for July 17, 2013. A curtailemnt of $380\si{\kilo\watt}$ is sustained during the DR event period.}
\label{fig:synthesis}
\vspace{-10pt}
\end{figure}

We now evaluate the performance of the mbCRT (Section~\ref{sec:mbcrt}) algorithm for real-time DR synthesis. 
Similar to DR evaluation, the regression tree is trained on weather, proxy features, set-point schedules and data from the building. 
We first partition the set of features into manipulated features (or control inputs) and non-manipulated features (or disturbances). 
There are three control inputs to the system: the chilled water set-point, zone air temperature set-point and lighting levels.
At design time, the model based tree built (Algorithm~\ref{alg:mbcrt}) has $369$ leaves and each of them has a linear regression model fitted over the control inputs with  the response variable being the power consumption of the building.

In addition to learning the power consumption prediction tree, $19$ additional model based trees were also built for predicting the different zone temperatures inside the building.
When the DR event commences, at every time-step (every 5 mins), DR-Advisor uses the mbCRT algorithm to determine which leaf, and therefore, which linear regression model will be used for that time-step to solve the linear program (Eq~\ref{eq:synth_program}) and determine the optimal values of the control inputs to meet a sustained response while maintaining thermal comfort.
 \begin{figure}
\centering
\includegraphics[width=\columnwidth]{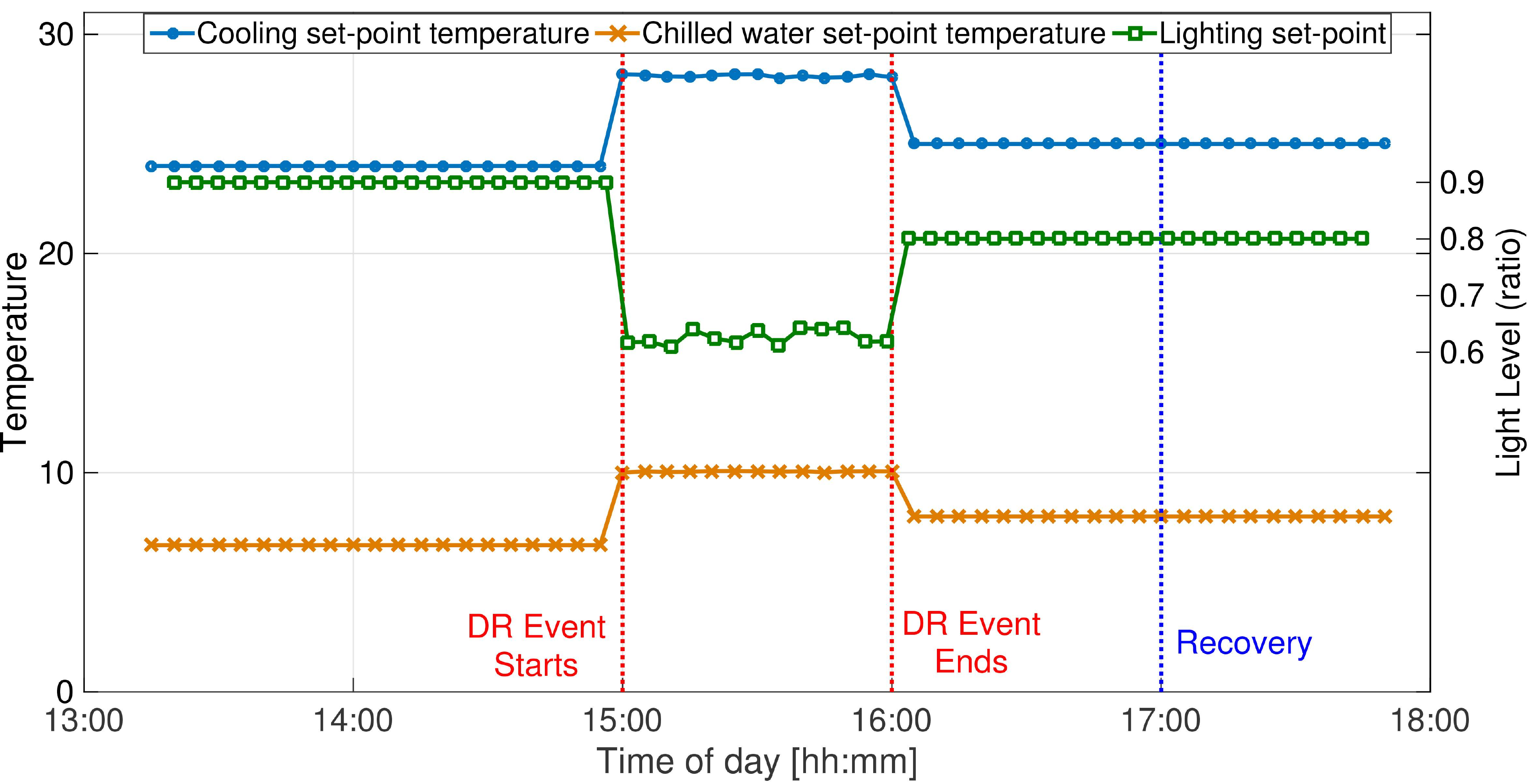}
\caption{Optimal DR strategy as determined by the mbCRT algorithm.}
\label{fig:set-points}
\vspace{-10pt}
\end{figure}

\begin{figure}
\centering
\includegraphics[width=\columnwidth]{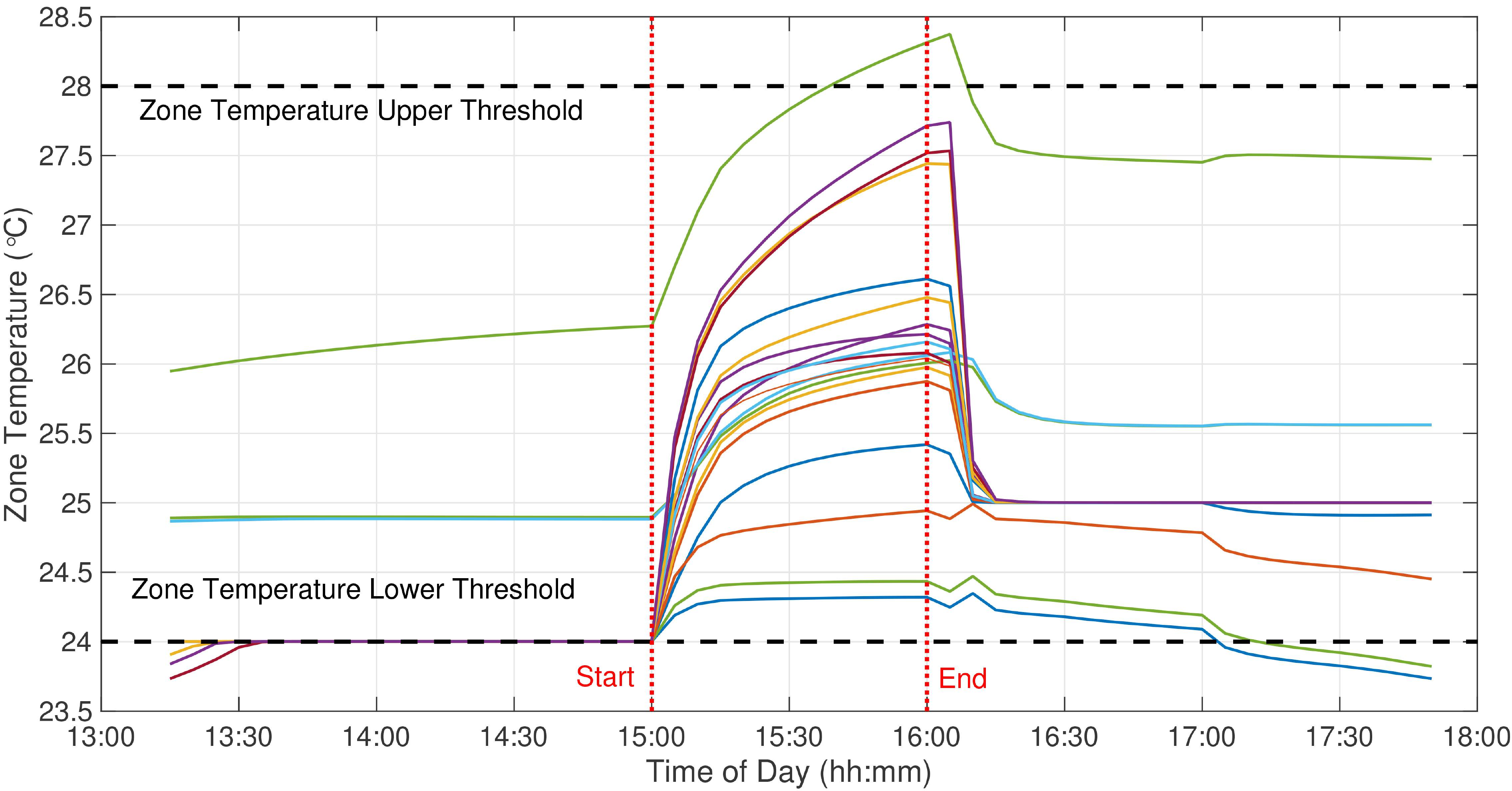}
\caption{The mbCRT algorithm maintains the zone temperatures within the specified comfort bounds during the DR event.}
\label{fig:alltemps}
\vspace{-10pt}
\end{figure}

\begin{figure}[b]
\includegraphics[width=\columnwidth]{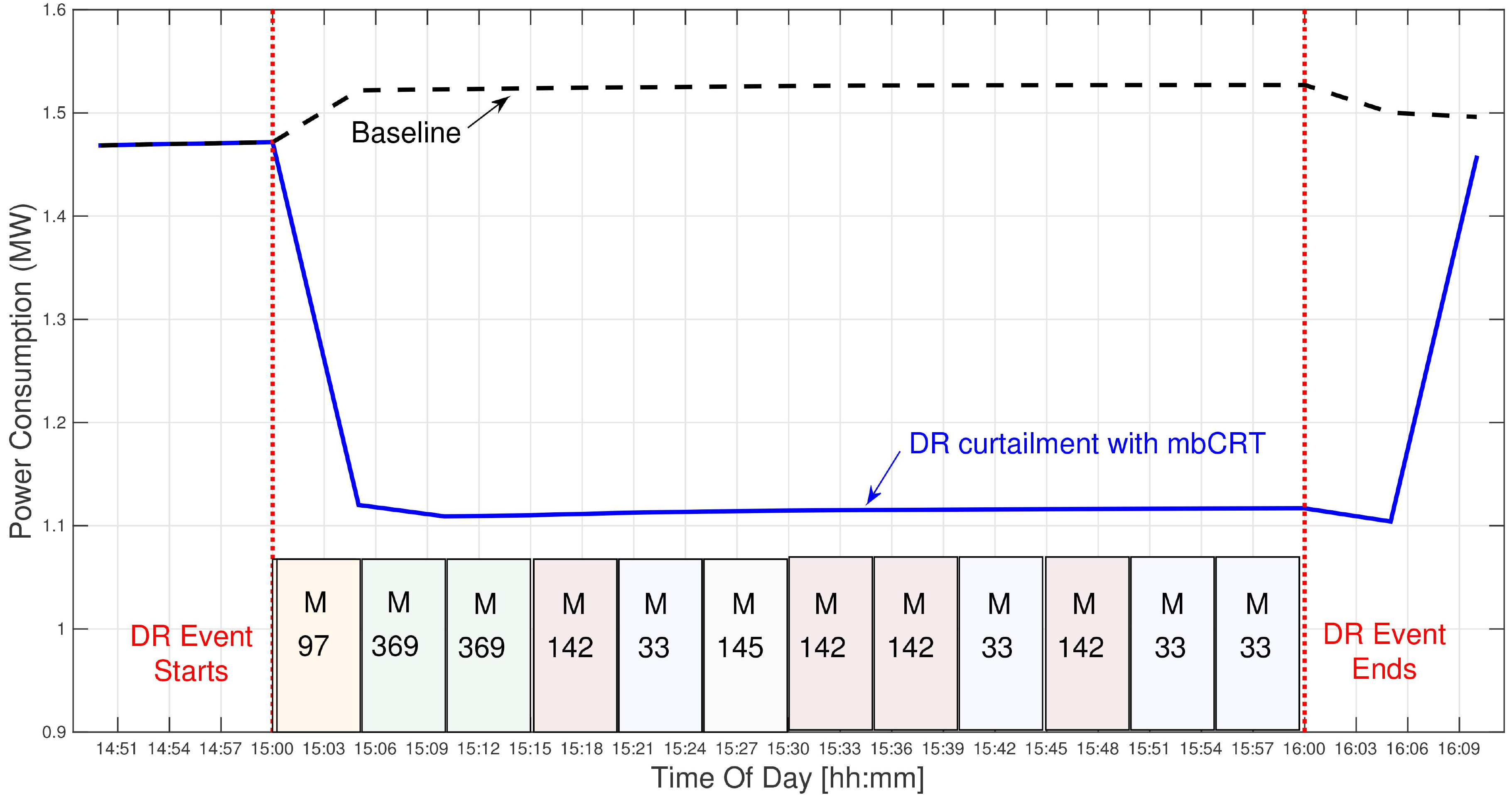}
\caption{Zoomed in view of the DR synthesis showing how the mbCRT algorithm selects the appropriate linear model for each time-step based on the forecast of the disturbances.}
\label{fig:model-sel}
\vspace{-10pt}
\end{figure}

Figure~\ref{fig:synthesis} shows the power consumption profile of the building using DR-Advisor for the DR event. 
We can see that using the mbCRT algorithm we are able to achieve a sustained curtailed response of $380\si{\kilo\watt}$ over a period of 1 hour as compared to the baseline power consumption estimate.  Also shown in the figure is the comparison between the best rule based fixed strategy which leads to the most curtailment in Section~\ref{sec:case_eval}. In this case the DR strategy synthesis outperforms the best rule base strategy (from Section~\ref{sec:case_eval}, Fig.~\ref{fig:case_eval_power}) by achieving a $17\%$ higher curtailment while maintaining thermal comfort. The rule-based strategy does not directly account for any effect on thermal comfort.
The DR strategy synthesized by DR-Advisor is shown in Figure~\ref{fig:set-points}. 
We can see in Figure~\ref{fig:alltemps} how the mbCRT algorithm is able to maintain the zone temperatures inside the building within the specified comfort bounds.
These results demonstrate the benefit of synthesizing optimal DR strategies as opposed ot relying on fixed rules and pre-determined strategies which do not account for any guarantees on thermal comfort. 
Figure~\ref{fig:model-sel} shows a close of view of the curtailed response. The leaf node which is being used for the power consumption constraint at every time-step is also shown in the plot.
We can see that the model switches several times during the event, based on the forecast of disturbances. 

These results show the effectiveness of the mbCRT algorithm to synthesize DR actions in real-time while utilizing a simple data-driven tree-based model.

\subsubsection{\textbf{Revenue from Demand Response}}
We use Con Edison utility company's commercial demand response tariff structure~\cite{edison} to estimate the financial reward obtained due to the curtailment achieved by the DR-Advisor for our Chicago based DoE commercial reference building.
The utility provides a $\$25/\si{\kilo\watt}$ per month as a reservation incentive to participate in the real-time DR program for summer. In addition to that, a payment of $\$1$ per kWh of energy curtailed is also paid. For our test-bed, the peak load curtailed is $380\si{\kilo\watt}$. If we consider $\sim5$ such events per month for 4 months, this amounts to a revenue of $\sim \$45,600$ for participating in DR which is $37.9\%$ of the energy bill of the building for the same duration ($\$120,317$).
This is a significant amount, especially since using DR-Advisor does not require an investment in building complex modeling or installing sensor retrofits to a building.

\section{Related Work}
\label{sec:related}
There is a vast amount of literature (\cite{auslanderdeep,oldewurtel2013towards,xu2004peak}) which addresses the problem of determining demand response strategies. 
The majority of approaches are using either rule-based approaches for curtailment or white/grey box model-based approaches.
These usually assume that the model of the system is either perfectly known or found in literature, whereas the task is much more complicated and time consuming in case of a real building and sometimes, it can be even more complex and involved than the controller design itself.
After several years of work on using first principles based models for demand response, multiple authors~\cite{sturzeneggermodel, vzavcekova2014towards} have concluded that the biggest hurdle to mass adoption of intelligent building control is the cost and effort required to capture accurate dynamical models of the buildings.
Since DR-Advisor learns an aggregate building level models and combined with the fact that weather forecasts are expected to become cheaper; there is little to no additional sensor cost of implementing the DR-Advisor recommendation system in large buildings. OpenADR standard and protocol~\cite{openadr} describes the formats for information exchange to facilitate DR but modeling, prediction and control strategies are out of scope.

Several machine learning approaches~\cite{edwards2012predicting, vaghefi2014modeling, yin2012scalable} have been utilized before for forecasting electricity load including some which use regression trees. 
However, there are three significant shortcomings of the work in this area: 
\begin{inparaenum}[(a)]
\item First, the time-scales at which the load forecasts are generated range from $15-20$ min upto an hour; which is too coarse grained for DR events which only last for at most a couple of hours and for real-time electricity prices which exhibit frequent changes. 
\item Secondly, these approaches are not aimed at solving demand response problems but are restricted to long term load forecasting with applications in evaluating building retrofits savings and building energy ratings. 
\item Lastly, in these methods, there is no focus on control synthesis or addressing the suitability of the model to be used in control design; whereas the mbCRT algorithm enables the use of regression trees for control synthesis with applications in demand response. 
\end{inparaenum}

\section{Conclusion}
\label{sec:discussion}
We present a data-driven approach for modeling and control of large scale cyber-physical energy systems which are inherently messy to model using first principles based methods.
We show how regression tree based methods are well suited to address challenges associated with demand response for large \textit{C/I/I} consumers while being simple and interpretable. 
We have incorporated all our methods into the DR-Advisor tool - \url{http://mlab.seas.upenn.edu/dr-advisor/}.

DR-Advisor achieves a prediction accuracy of $\textbf{92.8\%}$ to $\textbf{98.9\%}$ for eight buildings on the University of Pennsylvania's campus.
We compare the performance of DR-Advisor on a benchmarking data-set from AHRAE's energy predictor challenge and rank $2^nd$ among the winners of that competition.
We show how DR-Advisor can select the best rule-based DR strategy, which leads to the most amount of curtailment, from a set of several rule-based strategies. 
We presented a model based control with regression trees (mbCRT) algorithm which enables control synthesis using regression tree based structures for the first time. Using the mbCRT algorithm, DR-Advisor can achieve a sustained curtailment of $\textbf{380kW}$ during a DR event. Using a real tariff structure, we estimate a revenue of $\sim\$\textbf{45,600}$ for the DoE reference building over one summer which is $\textbf{37.9\%}$ of the summer energy bill for the building. The mbCRT algorithm outperforms even the best rule-based strategy by $\textbf{17\%}$.
DR-Advisor bypasses cost and time prohibitive process of building high fidelity models of buildings that use grey and white box modeling approaches while still being suitable for control design.
These advantages combined with the fact that the tree based methods achieve high prediction accuracy, make DR-Advisor an alluring tool for evaluating and planning DR curtailment responses for large scale cyber-physical energy systems.

\let\secfnt\undefined
\newfont{\secfnt}{ptmb8t at 10pt}

\bibliographystyle{unsrt}
\bibliography{references}

\end{document}